\documentclass[twocolumn]{webofc}
\usepackage[varg]{txfonts}   
\usepackage{colortbl} 
\usepackage{amsfonts}
\usepackage{amssymb}
\usepackage{amsmath}
\begin{document}
\title{The neutron electric dipole moment experiment at the Spallation Neutron Source}
%
%

\author{\firstname{K.K.H.} \lastname{Leung}\inst{1,2}\fnsep\thanks{\email{kkleung@ncsu.edu}} \and
\firstname{	M.	} \lastname{	Ahmed	}\inst{	2,3,4	}\and
\firstname{	R.	} \lastname{	Alarcon	}\inst{	5	}\and
\firstname{	A.	} \lastname{	Aleksandrova	}\inst{	6	}\and
\firstname{	S.	} \lastname{	Bae{\ss}ler	}\inst{	7,8	}\and
\firstname{	L.	} \lastname{	Barr{\' o}n-Palos	}\inst{	9	}\and
\firstname{	L.	} \lastname{	Bartoszek	}\inst{	10	}\and
\firstname{	D.H.	} \lastname{	Beck	}\inst{	11	}\and
\firstname{	M.	} \lastname{	Behzadipour	}\inst{	6	}\and
\firstname{	J.	} \lastname{	Bessuille	}\inst{	12	}\and
\firstname{	M.A.	} \lastname{	Blatnik	}\inst{	13	}\and
\firstname{	M.	} \lastname{	Broering	}\inst{	6	}\and
\firstname{	L.J.	} \lastname{	Broussard	}\inst{	8	}\and
\firstname{	M.	} \lastname{	Busch	}\inst{	2,3	}\and
\firstname{	R.	} \lastname{	Carr	}\inst{	13	}\and
\firstname{	P.-H.	} \lastname{	Chu	}\inst{	3	}\and
\firstname{	V.	} \lastname{	Cianciolo	}\inst{	8	}\and
\firstname{	S.M.	} \lastname{	Clayton	}\inst{	14	}\and
\firstname{	M.D.	} \lastname{	Cooper	}\inst{	14	}\and
\firstname{	C.	} \lastname{	Crawford	}\inst{	6	}\and
\firstname{	S.A.	} \lastname{	Currie	}\inst{	14	}\and
\firstname{	C.	} \lastname{	Daurer	}\inst{	11	}\and
\firstname{	R.	} \lastname{	Dipert	}\inst{	5	}\and
\firstname{	K.	} \lastname{	Dow	}\inst{	12	}\and
\firstname{	D.	} \lastname{	Dutta	}\inst{	15	}\and
\firstname{	Y.	} \lastname{	Efremenko	}\inst{	8,16	}\and
\firstname{	C.B.	} \lastname{	Erickson	}\inst{	11	}\and
\firstname{	B.W.	} \lastname{	Filippone	}\inst{	13	}\and
\firstname{	N.	} \lastname{	Fomin	}\inst{	16	}\and
\firstname{	H.	} \lastname{	Gao	}\inst{	3	}\and
\firstname{	R.	} \lastname{	Golub	}\inst{	1,2	}\and
\firstname{	C.R.	} \lastname{	Gould	}\inst{	1,2	}\and
\firstname{	G.L.	} \lastname{	Greene	}\inst{	8,16	}\and
\firstname{	D.G.	} \lastname{	Haase	}\inst{	1,2	}\and
\firstname{	D.	} \lastname{	Hasell	}\inst{	12	}\and
\firstname{	A.I.	} \lastname{	Hawari	}\inst{	1	}\and
\firstname{	M.E.	} \lastname{	Hayden	}\inst{	17	}\and
\firstname{	A.T.	} \lastname{	Holley	}\inst{	18	}\and
\firstname{	R.J.	} \lastname{	Holt	}\inst{	13	}\and
\firstname{	P.R.	} \lastname{	Huffman	}\inst{	1,2,8	}\and
\firstname{	E.	} \lastname{	Ihloff	}\inst{	12	}\and
\firstname{	T.M.	} \lastname{	Ito	}\inst{	14	}\and
\firstname{	J.	} \lastname{	Kelsey	}\inst{	12	}\and
\firstname{	Y.J.	} \lastname{	Kim	}\inst{	14	}\and
\firstname{	J.	} \lastname{	Koivuniemi	}\inst{	11	}\and
\firstname{	E.	} \lastname{	Korobkina	}\inst{	1,2	}\and
\firstname{	W.	} \lastname{	Korsch	}\inst{	6	}\and
\firstname{	S.K.	} \lastname{	Lamoreaux	}\inst{	19	}\and
\firstname{	E.	} \lastname{	Leggett	}\inst{	15	}\and
\firstname{	A.	} \lastname{	Lipman	}\inst{	1,2	}\and
\firstname{	C.-Y.	} \lastname{	Liu	}\inst{	20	}\and
\firstname{	J.	} \lastname{	Long	}\inst{	20	}\and
\firstname{	S.W.T.	} \lastname{	MacDonald	}\inst{	14	}\and
\firstname{	M.	} \lastname{	Makela	}\inst{	14	}\and
\firstname{	A.	} \lastname{	Matlashov	}\inst{	14}\thanks{Present address: IBS Center for Axion and Precision Physics Research}	\and
\firstname{	J.	} \lastname{	Maxwell	}\inst{	12	}\and
\firstname{	M.	} \lastname{	McCrea	}\inst{	6	}\and
\firstname{	M.	} \lastname{	Mendenhall	}\inst{	13	}\and
\firstname{	H.O.	} \lastname{	Meyer	}\inst{	20	}\and
\firstname{	R.	} \lastname{	Milner	}\inst{	12	}\and
\firstname{	P.	} \lastname{	Mueller	}\inst{	8	}\and
\firstname{	N.	} \lastname{	Nouri	}\inst{	6,19	}\and
\firstname{	C.M.	} \lastname{	O'Shaughnessy	}\inst{	14	}\and
\firstname{	C.	} \lastname{	Osthelder	}\inst{	13	}\and
\firstname{	J.-C.	} \lastname{	Peng	}\inst{	11	}\and
\firstname{	S.	} \lastname{	Penttila	}\inst{	8	}\and
\firstname{	N.S.	} \lastname{	Phan	}\inst{	14	}\and
\firstname{	B.	} \lastname{	Plaster	}\inst{	6	}\and
\firstname{	J.	} \lastname{	Ramsey	}\inst{	8,14	}\and
\firstname{	T.	} \lastname{	Rao	}\inst{	11	}\and
\firstname{	R. P.	} \lastname{	Redwine	}\inst{	12	}\and
\firstname{	A.	} \lastname{	Reid	}\inst{	1,2	}\and
\firstname{	A.	} \lastname{	Saftah	}\inst{	6	}\and
\firstname{	G.M.	} \lastname{	Seidel	}\inst{	21	}\and
\firstname{	I.F.	} \lastname{	Silvera	}\inst{	22	}\and
\firstname{	S.	} \lastname{	Slutsky	}\inst{	13	}\and
\firstname{	E.	} \lastname{	Smith	}\inst{	14	}\and
\firstname{	W.M.	} \lastname{	Snow	}\inst{	20	}\and
\firstname{	W.	} \lastname{	Sondheim	}\inst{	14	}\and
\firstname{	S.	} \lastname{	Sosothikul	}\inst{	1,2	}\and
\firstname{	T.D.S.	} \lastname{	Stanislaus	}\inst{	23	}\and
\firstname{	X.	} \lastname{	Sun	}\inst{	13	}\and
\firstname{	C.M.	} \lastname{	Swank	}\inst{	13	}\and
\firstname{	Z.	} \lastname{	Tang	}\inst{	14	}\and
\firstname{	R.	} \lastname{	Tavakoli Dinani	}\inst{	17	}\and
\firstname{	E.	} \lastname{	Tsentalovich	}\inst{	12	}\and
\firstname{	C.	} \lastname{	Vidal	}\inst{	12	}\and
\firstname{	W.	} \lastname{	Wei	}\inst{	13,14	}\and
\firstname{	C.R.	} \lastname{	White	}\inst{	1,2	}\and
\firstname{	S.E.	} \lastname{	Williamson	}\inst{	11	}\and
\firstname{	L.	} \lastname{	Yang	}\inst{	11	}\and
\firstname{	W.	} \lastname{	Yao	}\inst{	8	}\and
\firstname{	A.R.	} \lastname{	Young	}\inst{	1,2	}
}
\institute{
North Carolina State University, Raleigh, NC 27695, USA \and 
Triangle Universities Nuclear Laboratory, Durham, NC 27708, USA \and 
Duke University, Durham, NC 27708, USA \and 
North Carolina Central University, Durham, NC 27707, USA \and 
Arizona State University, Tempe, AZ 85287, USA  \and 
University of Kentucky, Lexington, KY 40506, USA  \and 
University of Virginia, Charlottesville, VA 22904, USA \and 
Oak Ridge National Laboratory, Oak Ridge, TN 37831, USA \and 
Universidad Nacional Auton{\' o}ma de Mexico, Apartado Postal 20-364, M{\'e}xico \and 
Bartoszek Engineering, Aurora, IL 60506, USA  \and 
University of Illinois Urbana-Champaign, Champaign, IL 61801, USA  \and 
Massachusetts Institute of Technology, Cambridge, MA 02139, USA  \and 
California Institute of Technology, Pasadena, CA 91125, USA  \and 
Los Alamos National Laboratory, Los Alamos, NM 87545, USA \and 
Mississippi State University, Mississippi State, MS 39762, USA \and 
University of Tennessee, Knoxville, TN 37996, USA \and 
Simon Fraser University, Burnaby, BC, V5A 1S6, Canada \and 
Tennessee Technological Institute, Cookeville, TN 38501, USA \and 
Yale University, New Haven, CT 06520, USA \and 
Indiana University, Bloomington, IN 47408, USA \and 
Brown University, Providence, RI 02912, USA \and 
Harvard University, Cambridge, MA 02138, USA \and 
Valparaiso University, Valparaiso, IN 46383, USA 
}
          
\abstract{Novel experimental techniques are required to make the next big leap in neutron electric dipole moment experimental sensitivity, both in terms of statistics and systematic error control. The nEDM experiment at the Spallation Neutron Source (nEDM@SNS) will implement the scheme of Golub \& Lamoreaux [Phys. Rep., 237, 1 (1994)]. The unique properties of combining polarized ultracold neutrons, polarized $^3$He, and superfluid $^4$He will be exploited to provide a sensitivity to $\sim 10^{-28}\,e{\rm \,\cdot\, cm}$. Our cryogenic apparatus will deploy two small ($3\,{\rm L}$) measurement cells with a high density of ultracold neutrons produced and spin analyzed in situ. The electric field strength, precession time, magnetic shielding, and detected UCN number will all be enhanced compared to previous room temperature Ramsey measurements. Our $^3$He co-magnetometer offers unique control of systematic effects, in particular the Bloch-Siegert induced false EDM. Furthermore, there will be two distinct measurement modes: free precession and dressed spin. This will provide an important self-check of our results. Following five years of ``critical component demonstration,'' our collaboration transitioned to a ``large scale integration'' phase in 2018. An overview of our measurement techniques, experimental design, and brief updates are described in these proceedings.}
\maketitle
\section{Introduction}
\label{intro}
The existence of a permanent neutron electric dipole moment $d_n$ would be a direct violation of time-reversal ($T$) symmetry. This is equivalent to violation of combined charge and parity ($CP$) symmetry via the $CPT$ theorem \cite{Schwinger1951,Luders1957}, which has so far withstood experimental scrutiny. $CP$ violation has been measured in the $K$ meson and $B$ meson systems \cite{Christenson1964,Aubert2002,Abe2002} and is successfully incorporated in the standard model as a complex phase in the Cabibbo-Kobayashi-Maskawa quark mixing matrix \cite{Cabibbo1963,Kobayashi1973}. While $CP$ violation is a key ingredient to generate a baryon-antibaryon asymmetry \cite{Sakharov1967}, the amount of known standard model $CP$ violation is much too small to adequately explain the observed size of this asymmetry in the Universe.

In electroweak baryogenesis mechanisms proposed by minimal supersymmetric extensions to the standard model, the lower limit on the size of the neutron electric dipole moment (EDM) required is $|d_n| > \mathcal{O}(10^{-28}\, e{\rm \,\cdot\, cm})$ to be consistent with the baryon-to-photon ratio measured from the Cosmic Microwave Background \cite{Li2009,Cirigliano2010}. This is significantly larger than the value $|d_n| \sim \mathcal{O}(10^{-31}\, e{\rm \,\cdot\, cm})$ inferred from the standard model. Therefore, searching for a neutron EDM down to the $\mathcal{O}(10^{-28}\, e{\rm \,\cdot\, cm})$ level provides a promising route for the discovery of new physics. If an observation was made at this level, it would represent a clear signal above the standard model calculated ``background''. Even a negative result would have significant impact as it would rule out a large class of beyond standard model theories.


In 1950 the first neutron EDM experiment was performed using Ramsey's separated oscillatory fields technique \cite{Ramsey1950} on a neutron beam in an applied electric field \cite{Smith1957}. The refinement of this method over the next three decades \cite{Ramsey1982} improved its sensitivity by four orders-of-magnitude \cite{Dress1977}. Since the 1980s, a transition to using stored ultracold neutrons (UCNs) and the implementation of a cohabiting magnetometer species (``co-magnetometer'') to correct for temporal and spatial magnetic field inhomogeneities (while still using Ramsey's technique) were used to attain a sensitivity improvement of a further two orders-of-magnitude \cite{Lamoreaux2009}. The current world limit is $|d_n| < 3.0 \times 10^{-26}\,e{\rm \,\cdot\, cm}$ (90~\% C.L.) \cite{Baker2006,Pendlebury2015}. In this experiment, UCNs from the PF2 turbine source at the Institut Laue-Langevin (ILL) were used to fill a 21~L cylindrical measurement cell (with a $47\,{\rm cm}$ diameter) to reach an initial polarized density of $\approx 3.5\,{\rm UCN\,cm^{-3}}$ \cite{Baker2014}. An electric field of $E \approx 10\,{\rm kV/cm}$ was used, along with a Ramsey free precession time $T_m\approx 130\,{\rm s}$. The choice of $T_m$ is determined by a combination of the UCN storage time and spin coherence times of the UCNs and the co-magnetometer. The detected UCN counts after $T_m$ corresponded to a density in the cell of $\approx 0.7\,{\rm UCN\,cm^{-3}}$ with a polarization observable (i.e. the contrast of the Ramsey fringes) of $\alpha \approx 0.6$.

Relocation of this apparatus from the ILL to a new solid deuterium source at the Paul Scherrer Institute \cite{Anghel2009,Becker2015} gave a 10\% increase in the UCN numbers, but more importantly, improvements in other systems allowed $E\rightarrow15\,{\rm kV/cm}$, $T_{m}\rightarrow 180\,{\rm s}$, and $\alpha \rightarrow 0.8$ \cite{Roccia2018}. The shot noise statistical sensitivity of a neutron EDM experimental cycle is given by:
\begin{equation}
\sigma(d_n) = \frac{\hbar}{2\alpha E T_{m} \sqrt{N}}\;,
\end{equation}
where $N$ is the number of detected UCNs. Therefore, the sensitivity of the experiment per day improved by more than a factor of three.

There is ample room for increasing the number of UCNs in future neutron EDM experiments. However, there exists a Bloch-Siegert shift induced false EDM \cite{Commins1991,Pendlebury2004} (often called the ``geometric-phase induced frequency shift'') caused by the interaction of the motion-induced ${\bf E} \times{\bf v}/c^2$ field with magnetic field gradients, where ${\bf v}$ is the velocity. For a spin species undergoing diffusive motion (e.g. the co-magnetometer), the size of this false EDM is proportional to $L^2$, where $L$ is the long dimension of the cell. Furthermore, the transverse spin coherence time scales as $L^{-4}$ \cite{McGregor1990}. Therefore, there are limits to the extent that $N$ can be increased by increasing the size of the cell alone.

To reach a $\mathcal{O}(10^{-28}\, e{\rm \,\cdot\, cm})$ sensitivity new techniques are needed. A novel experimental scheme that can reach this coveted precision in both statistics and systematics was proposed in \cite{Golub1994a}. Here, unique properties of polarized $^3$He and superfluid $^4$He are exploited to allow a neutron EDM experiment to be performed on a high density of polarized UCNs located inside two small $\sim 3\,{\rm L}$ measurement cells. This experiment, which we refer to as nEDM@SNS, will be located at the Spallation Neutron Source at Oak Ridge National Laboratory. A more in-depth description of our experiment will be published shortly \cite{Filippone2019}.

\section{nEDM@SNS experimental scheme}

\begin{figure}
\centering
\includegraphics[width=\columnwidth,clip]{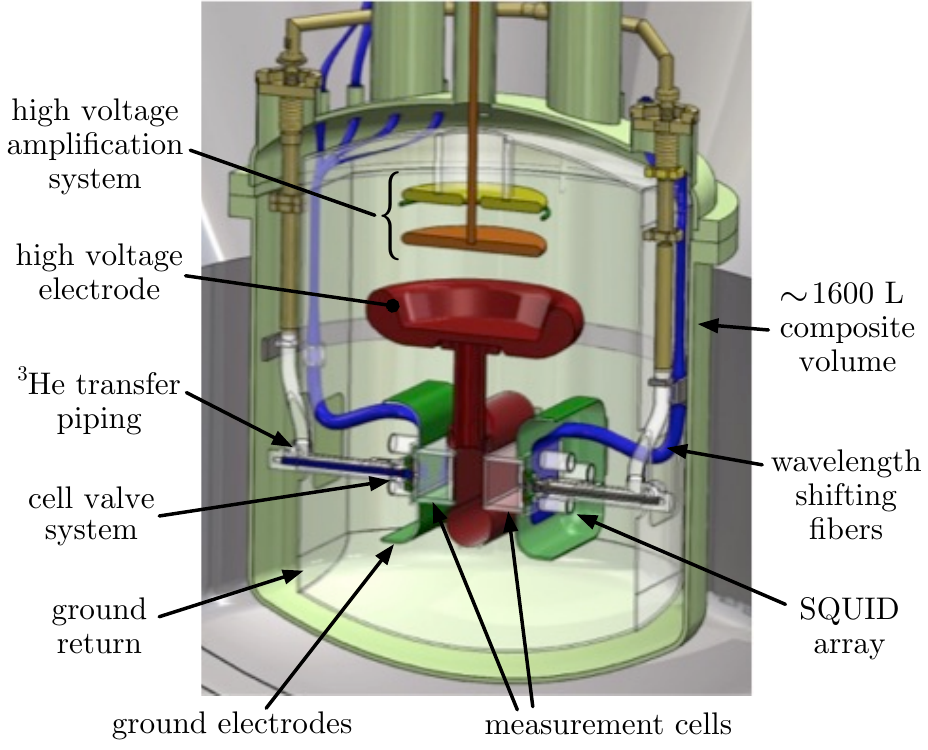}
\caption{The design of the Central Detector System. The ${\sim 1600\,{\rm L}}$ volume, made from composite materials, is filled with superfluid $^4$He cooled to $\sim 0.4 \,{\rm K}$ with a dilution refrigerator (not shown). This system contains the two measurement cells filled with isotopically pure superfluid $^4$He along with systems for the electric field, light collection, $^3$He cell valve, and SQUIDs. The cold neutron beam goes into the page. For scale, the inner cell dimensions are $7.5\,{\rm cm} \,(W) \times 10\,{\rm cm}\,(H) \times 40\,{\rm cm}\,(L)$ relative to the beam.}
\label{fig:CDS} 
\end{figure}

At the heart of the apparatus will be two measurement cells both $40\,{\rm cm}$ in length and with cross-sections $7.5\,{\rm cm}\, (W) \times 10\,{\rm cm}\, (H)$ filled with isotopically pure superfluid $^4$He cooled to $\sim 0.4\,{\rm K}$. Along the short horizontal dimension of both cells, a highly homogenous magnetic field $B_0 = 30\,{\rm mG}$ will be applied, a value chosen to balance between expected systematic uncertainties and statistical precision. Polarized $^3$He will be preloaded into the cells at an isotopic concentration of $x_3 \sim 10^{-10}$ and a polarization $P_3\sim 0.98$. A high voltage electrode will be located between the two cells, while two ground electrodes will be situated on either side (see Fig.~\ref{fig:CDS}). Therefore, ${\bf E}$ will be parallel to ${{\bf B}_0}$ in one cell and anti-parallel to it in the other.

The FNPB cold neutron beam at the SNS \cite{Fomin2015}, collimated and focused to be smaller than the cells' inner dimensions, will be turned on to produce and ``fill'' UCNs inside the cell \cite{Golub1975,Golub1977}. This we call \emph{in situ} UCN production. The initially white beam will be polarized using a super-mirror and will be chopped to $8.9\AA \pm 1\%$ to suppress backgrounds caused by cold neutrons with wavelengths other than at the single-phonon UCN production peak. On the inside of the cell there will be a deuterated polystyrene (dPS) coating, which has a neutron optical potential $V_{\rm dPS} \approx 160\,{\rm neV}$. The volumetric production rate of polarized UCNs that can be stored by this potential will be $P_{\rm UCN/V} \approx 0.31\,{\rm UCN\,cm^{-3}\,s^{-1}}$.

The UCN density in the cell will build-up with a time constant $\tau_{\rm fill}$ given by: $\tau_{\rm fill}^{-1} = \tau^{-1}_{\rm \beta} + \tau^{-1}_{\rm up} +  \tau^{-1}_{\rm walls} + \tau^{-1}_{3}$. These time constants are: the neutron $\beta$-decay lifetime $\tau_{\rm \beta} \approx 880\,{\rm s}$; the UCN upscattering loss lifetime set by interactions with the superfluid helium excitations, expected to be $\tau_{\rm up} \approx 7\times 10^{4}\,{\rm s}$ at $0.4\,{\rm K}$ \cite{Golub1979,Golub1983}; the UCN wall loss lifetime at the cell walls, where the design goal is $\tau_{\rm walls} > 2,000\,{\rm s}$; and the UCN-$^3$He capture lifetime, which during filling will be $\tau_{\rm 3} \gtrsim 10^{4}\,{\rm s}$. Combining these time constants and assuming good match between the neutron beam dimensions with the cell, the saturated UCN density in the cell will thus be $\approx 170\,{\rm UCN\,cm^{-3}}$. The UCN polarization is expected to be high, $P_n \approx 0.98$, since the cold neutrons retain their polarization after scattering off superfluid $^4$He phonons, and the UCNs in the opposite spin-state will be absorbed by the $^3$He during accumulation with a time constant of  approximately 200\,{\rm s}.

With in situ UCN production, and also having the UCN spins analyzed in the cell (which we call ``\emph{in situ} spin analysis,'' described next in Sec.~\ref{sec:spinanalysis}), the combined $\sim 1\times10^{6}$ UCNs produced in the two cells can be utilized. This avoids UCN loss and depolarization that can occur during transport from an external UCN source and then to an external spin analysis system. These problems have a long history of degrading the expected performance of UCN experiments. The physics of the super-thermal UCN production process in superfluid helium has been well-studied \cite{Golub1983,Yoshiki1992,Baker2003,Zimmer2007a,Piegsa2014,Leung2016}.

Our in situ UCN production and in situ spin analysis techniques provide a large statistical sensitivity increase compared to neutron EDM experiments that use external UCN sources because of the increase in detected UCN number $N$. The other statistical advantages of our experimental design include the suppression of UCN upscattering loss from the cryogenic cell walls and the large electric field $|E| \gtrsim 75\,{\rm kV\,cm^{-1}}$ \cite{Gerhold1998,Ito2016} that can be supported by the superfluid helium due to its high dielectric breakdown strength.

For the suppression and control of systematic effects, our cryogenic environment allows the natural implementation of superconducting magnetic shielding, which is particularly useful for suppressing low frequency ambient field drifts. Furthermore, by making small changes to the temperature of the superfluid, the mean free path of the $^3$He co-magnetometer can be changed drastically. This has important consequences for controlling and suppressing important systematic errors, as will be described in Sec.~\ref{sec:3Hecomagnetometer}. First, the role of the $^3$He as UCN spin analyzer will be described.

\subsection{UCN spin analysis with polarized $^3$He \label{sec:spinanalysis}}

UCNs experience the strongly spin-dependent capture reaction ${\rm n + {^3}He \rightarrow p + {^3}H + 764\,keV}$. For anti-parallel neutron and $^3$He nuclear spins, the capture cross-section at thermal energies is $\approx 11\,{\rm kb}$. This translates to a cross-section of $\approx 800\,{\rm kb}$ at a relative velocity of $\approx 30\,{\rm m\,s^{-1}}$ in our experiment. When the spins are parallel, the absorption cross-section is small; the experimental limit on the triplet to singlet absorption cross-section ratio is $\lesssim 1\%$ \cite{Passell1966}.

After a ${\rm n + {^3}He}$ capture event, the charged products will produce scintillation light in the superfluid helium at wavelengths of $\sim 80\,{\rm nm}$. This EUV light inside the cell will be converted to blue light using deuterated 1,1,4,4-tetraphenyl-1,3-butadiene (dTPB) that is embedded in the dPS polymer matrix on the inner cell walls. Since the UCNs do not leave the measurement cell, the analysis of their spin orientation is performed in situ. The detection of this scintillation light will also give live information on the average relative angle between the UCN and $^3$He spins throughout a measurement. This is different than the Ramsey technique where the phase of the UCN spins (converted to a longitudinal polarization) is only measured after the end of the precession period. This opens up possibilities for different measurement schemes.

Quantitatively, the average scintillation light event rate will be given by:
\begin{equation}
\Phi(t) = N(t) \biggl\{ \frac{\epsilon_\beta}{\tau_\beta} +\frac{\epsilon_3}{\bar{\tau}_3} \Big[ 1 - P_3(t) P_n(t) \cos\theta_{3n}(t) \Big] \biggr\} + R_{\rm BG}\, ,
\label{eq:scintillationRate}
\end{equation}
where $N(t)$ is the number of UCNs remaining in the cell at time $t$, $P_3(t)$ and $P_n(t)$ are the $^3$He and UCN polarizations, $\theta_{3n}(t) \equiv \theta_3(t) - \theta_n(t)$ is the difference between the average phase angle of the two spin species, and $\bar{\tau}_3 \approx (3.9\times10^{-8}\,{\rm s})x_3^{-1} $ is the time-averaged n-$^3$He absorption time constant.

UCN $\beta$-decay, ${\rm n \rightarrow p + e^- + \bar{\nu}}$, produces a time-dependent ``background'' in $\Phi(t)$. The scintillation light produced from this reaction comes primarily from the electron, which has an energy between $0$ and $783\,{\rm keV}$. Due to a combination of the stopping power differences between electrons, protons and tritons, and the effects of the electric field, the n-$^3$He capture peak in the number of EUV photon spectrum is expected to sit on top of a broad $\beta$-decay bump \cite{Ito2012,Ito2013}. By making appropriate cuts in the detected spectrum, a proportion of the $\beta$-decay events can be rejected \cite{Archibald2006}. For our scintillation light detection system design (described in Sec.~\ref{sec:expDesign}), the statistically optimized acceptance probabilities in Eq.~\ref{eq:scintillationRate} are $\epsilon_3 \approx 0.93$ and $\epsilon_\beta \approx 0.33$ for ${\rm n + {^3}He}$ capture events and neutron $\beta$-decay, respectively. The rate of ambient background events that will fall inside the spectrum cuts are included as $R_{\rm BG}$ in Eq.~\ref{eq:scintillationRate}.

\subsection{Free precession measurement mode \label{sec:freeprecess}}

Our experiment can be performed employing two distinct measurement modes. In this section, we describe the free precession mode. (The dressed spin mode is described in Sec.~\ref{sec:dressedspin}.) In both modes the measurements start after the cell is filled with UCNs from the cold neutron beam followed by a $\pi/2$ pulse, an applied AC magnetic field that rotates both the $^3$He and UCN spins into the plane transverse to ${\bf B}_0$.

In the free precession mode, the two spin species will simply be left to freely precess after the $\pi/2$ pulse. Due to the interaction with the combined magnetic and electric fields, $\theta_{3n}(t)$ evolves as:
\begin{equation}
\theta_{3n}(t) = \left[(\gamma_n-\gamma_3) B_0 \pm \frac{2 d_n |E|}{\hbar} \right]t + \phi_0 \equiv \omega_{3n}^\pm t +\phi_0 \;,
\label{eq:basicOmegaFree}
\end{equation}
where $\gamma_n$ and $\gamma_3$ are the neutron and $^3$He gyromagnetic ratios, $\phi_0$ is the phase at the start of the measurement, and $\omega^\pm_{3n}$ is the angular precession frequency difference between the neutron and $^3$He. The `$+$' or `$-$' signs and superscripts are for ${\bf E}$ parallel or anti-parallel to ${\bf B}_0$, respectively. For $B_0 = 30\,{\rm mG}$, since $\gamma_3 \approx -2.0 \times 10^8\,{\rm rad\,s^{-1}\,T^{-1}}$ and $\gamma_3/\gamma_n\approx 1.1$, $f_{3n} \equiv \omega_{3n}/(2\pi) \approx -9.8 \,{\rm Hz}$. A negative frequency corresponds to clockwise motion looking down the ${\bf B}_0$ axis. Of course, the $E$ field would also interact with any $^3$He EDM, but the size of this effect will be greatly suppressed due to screening by atomic electrons \cite{Schiff1963,Flambaum2012}.


If we include transverse spin relaxation of the UCNs and $^3$He, then $P_n(t) = P_n^0 \exp(-t/T_{\rm 2,n})$ and $P_3(t) = P_3^0 \exp(-t/T_{\rm 2,3})$. Our design goals for both $T_{\rm 2,n}$ and $T_{\rm 2,3}$ are $2\times 10^4\,{\rm s}$. The scintillation light event rate in the free precession mode will thus be given by:
\begin{align}
\begin{split}
\Phi(t) = N(t) \biggl\{ \frac{\epsilon_\beta}{\tau_\beta} +\frac{\epsilon_3}{\bar{\tau}_3}  \Big[ 1 - &P_n^0 P_3^0{\rm e}^{-(T^{-1}_{\rm 2,n}+T^{-1}_{\rm 2,3})t} \\*
&\times  \cos(\omega_{3n} t + \phi_0) \Big] \biggr\} + R_{\rm BG}. 
\label{eq:scintillationRate2}
\end{split}
\end{align}
Defining $\bar{\tau}_{\rm tot}^{-1} =  \tau_\beta^{-1} + \bar{\tau}_{\rm walls}^{-1} + \bar{\tau}_3^{-1}$, the number of UCNs in the cell will be given by:
\begin{align}
\begin{split}
N(t) = N_0 \exp \Bigg[-\frac{t}{\bar{\tau}_{\rm tot}} + \frac{P^0_n  P^0_3}{\bar{\tau}_3} & \int_0^t {\rm e}^{-(T^{-1}_{\rm 2,n}+T^{-1}_{\rm 2,3})t'} \\*
 \times & \cos(\omega_{3n} t' + \phi_0) \,{\rm d}t' \Bigg]\;,
\label{eq:numUCN}
\end{split}
\end{align}
where $N_0$ is the initial number of UCNs. The integral produces small short-lived oscillations in $N(t)$ that for the level of description in these proceedings can be neglected so that:
\begin{equation}
N(t) \approx N_0 \exp( -t/\bar{\tau}_{\rm tot})\;.
\end{equation}

In the absence of systematic effects, the difference between two $\omega_{3n}$ measurements when ${\bf E}$ is reversed relative to ${\bf B}_0$ would be given by:
\begin{equation}
\Delta \omega_{3n} \equiv \omega_{3n}^{+} - \omega_{3n}^{-} =  \frac{4 d_n |E|}{\hbar} \;.
\label{eq:freqDiffEDMsimple}
\end{equation}
To search for a non-zero neutron EDM signal, the difference between $\Delta \omega_{3n}$ extracted from simultaneous measurements in the two cells (i.e. made at different spatial positions) and from measurements as the polarity on the high voltage electrode alternated (i.e. made at different times) will be analyzed. 

\begin{figure}
\centering
\includegraphics[width=7.5cm,clip]{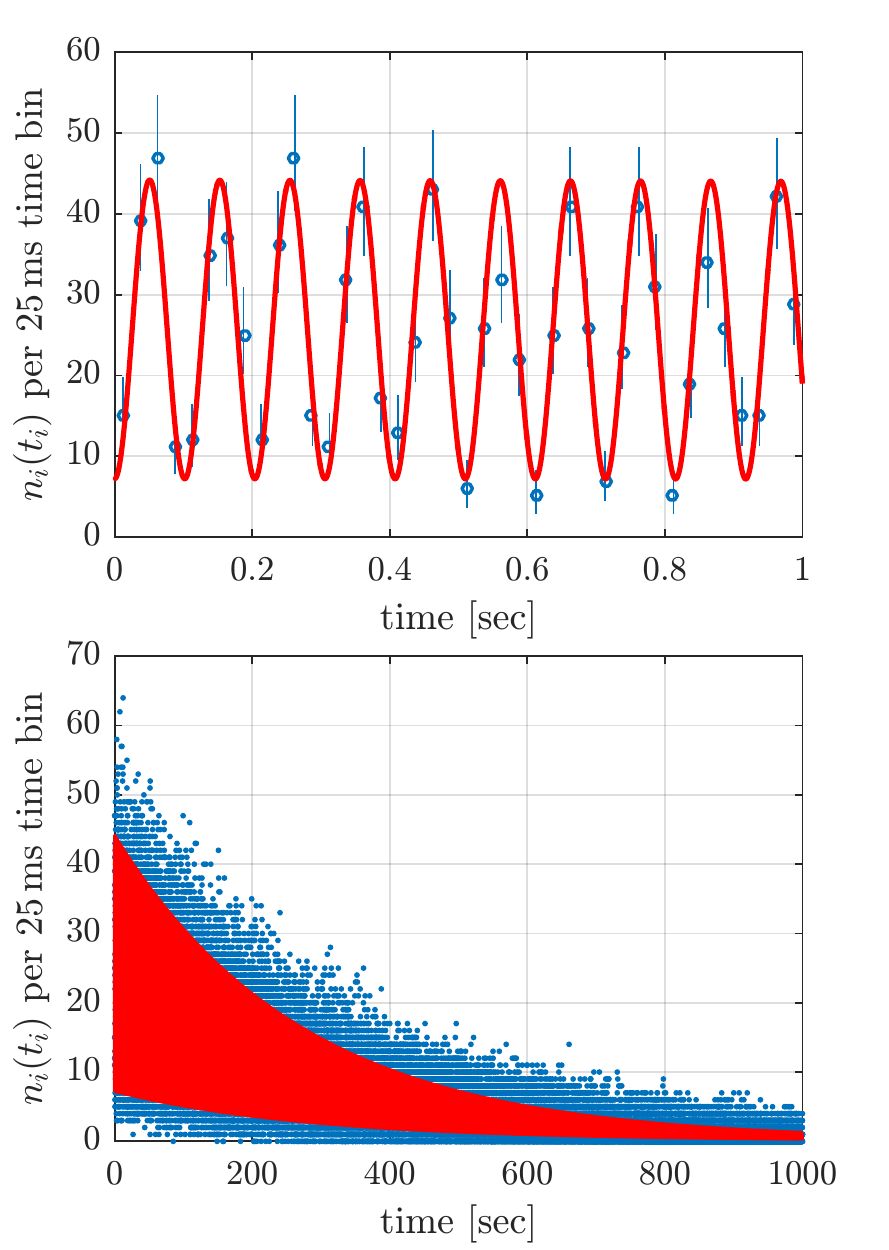}
\caption{Example of a Monte Carlo simulation of the scintillation event rate vs time data during the free precession period of a single free precession measurement cycle. The events are time binned, with $n_i(t_i)$ being the number of events that fall within a time bin centered at time $t_i$. The parameters used are described in the text. The best fit curve is shown in red. The upper plot is zoomed in on the first 1\,{\rm s}, with the error bars shown as the 67\% C.I. for a Poisson distribution. The lower plot is of the entire period and without error bars on the points.}
\label{fig:freePrecess} 
\end{figure}

Monte-Carlo simulations of the scintillation event rate versus time during the free precession period have been made and fitted (see Fig.~\ref{fig:freePrecess}). The fit is performed over the whole free precession measurement time $T_m$. However, since we are able to observe the live UCN precession with our technique, the frequency analysis can be performed in short time windows. This will be useful for correcting temporal magnetic field drifts (see Sec.~\ref{sec:3Hecomagnetometer}).

At the end of every measurement cycle, the (partially) depolarized $^3$He will be removed from the cell and new polarized $^3$He injected into the cell (described in Sec.~\ref{sec:expDesign}). The ${\bf E}$ field may also be reversed systematically between measurement cycles. The combined time for these operations is the dead time. For the sensitivity discussions next, a dead time of $400\,{\rm s}$ is assumed, along with an ambient background rate of $R_{\rm BG} = 5\,{\rm s^{-1}}$. The influence of these parameters on the sensitivity have been studied. Additionally, $\phi_0$ is assumed to be fixed in the fitting routine for the discussions below. This requires $\phi_0$ to be reproducible after the $\pi/2$ pulse.

Our simulations along with the above assumptions lead to the conclusion that the statistically optimum operating values for extracting $d_n$ from repeated measurement cycles include: (a) a value of $x_3$ that produces $\bar{\tau}_3 \approx 500\,{\rm s}$ (i.e. $\bar{\tau}_{\rm tot} \approx 270\,{\rm s} $, assuming $\tau_{\rm walls} = 2000\,{\rm s}$), (b) $T_m \approx 1000\,{\rm s}$, and (c) a cold neutron beam fill time $T_{\rm fill} \approx 1000\,{\rm s}$. The $1\sigma$ uncertainty in extracting $f_{3n}$ from these operating parameters is $\sigma_{f_{3n}} = 1.7\,{\rm \mu Hz}$ per free precession cycle per cell. For an electric field $E = 75\,{\rm kV/cm}$, the $1\sigma$ precision in extracting a neutron EDM is $\sigma_{d_n} \approx 3\times 10^{-28}\,e{\rm \,\cdot\, cm}$ for 300 live days of running. We expect to be able to achieve this in around three calendar years.

Before describing the dressed spin measurement mode, we first discuss the role of the $^3$He as a co-magnetometer and the key systematic effects involved in the free precession mode.

\subsection{$^3$He co-magnetometer and systematic effects} \label{sec:3Hecomagnetometer}

When writing Eq.~(\ref{eq:freqDiffEDMsimple}), it was implied that the static $B_0$ field in the two $\omega_{3n}$ measurements are identical. However, since they are made for cells at different spatial positions or in the same cell during different times, corrections for magnetic field gradients and drifts will inevitably be required. Therefore, high-precision magnetometry is needed.

A magnetometer system located outside the measurement cell and away from the large electric field can be used to infer the magnetic field inside the cells. However, such external magnetometry systems are insensitive to fields produced by leakage currents \cite{Lamoreaux2009} or magnetization contamination \cite{Pendlebury2015} near the cell, both of which can produce sizable systematic effects. For high precision neutron EDM experiments, co-magnetometery is needed.


The polarized $^3$He located in the same volume as the UCNs can be used as a co-magnetometer. For a concentration of $x_3 = 10^{-10}$, the $^3$He number density is $\sim 10^{12}\,{\rm cm^{-3}}$ (compared with  $\lesssim 10^{3}\,{\rm cm^{-3}}$ for the UCN density). The precessing $^3$He magnetization will generate an oscillating field with an amplitude of several fT close to the cell. This will be measured by an array of Superconducting Quantum Interference Device (SQUID) magnetometers near the cell \cite{Kim2013}.

While co-magnetometry drastically reduces systematic effects due to field variations, it adds additional effects at a much smaller scale. These come in the form of different ensemble averaged magnetic fields between the co-magnetometer atoms and the UCNs, despite the fact they occupy the same volume, as well as different precession frequency shifts experienced by the two.

The observed $\omega_{3n}$ in a cell should be written as:
\begin{equation}
\omega_{3n}^\pm = \gamma_n \langle{B}_{n}\rangle - \delta{\omega_n} - \gamma_3 \langle{B}_{3}\rangle + \delta{\omega_3} \pm \frac{2 d_n |E|}{\hbar} \, ,
\label{eq:complexOmegaFreewFreqShifts}
\end{equation}
where $\langle{B}_{3}\rangle$ and $\langle{B}_{n}\rangle$ are the ensemble averaged magnetic fields, and $\delta{\omega_3}$ and $\delta{\omega_n}$ are the frequency shifts of the two species. These effects only arise because of the differences in motion and precession between the co-magnetometer atoms and UCNs.

Firstly, the $^3$He atoms are in thermal equilibrium with the superfluid helium bath. Their velocity, taking into account the $^3$He effective mass, will be $\approx 30\,{\rm m\,s^{-1}}$. The UCNs, however, will have velocities $\approx 4\,{\rm m\,s^{-1}}$. It is worth mentioning that this velocity difference will be smaller than for other commonly used co-magnetometers at room temperature. Secondly, UCNs undergo ballistic motion reflecting only off the cell walls whereas the $^3$He motion will be diffusive. At our design concentration the $^3$He mean-free-path is approximately given by ${0.77\,{\rm mm}} \times [(0.45\,{\rm K})/T]^{15/2}$ \cite{Baym2013}, dominated by scattering off phonons in the superfluid. Thirdly, as already mentioned, the spin precession frequency of the $^3$He will only be $\approx 10\%$ faster than the UCNs. Again, this will be a smaller difference than compared with typical co-magnetometers. Furthermore, it should be noted that the sign of $\gamma_3$ and $\gamma_n$ are both negative. This will suppress effects such as that caused by Earth's rotation \cite{Lamoreaux2009}.

The most serious of these frequency shifts that can produce a false EDM signal comes from an interaction between magnetic field gradients and the motional magnetic field ${\bf B}_{E \times v} = {\bf E} \times {\bf v}/c^2$ \cite{Lamoreaux2005,Barabanov2006, Golub2011, Swank2012a,Steyerl2014, Golub2015, Pignol2015, Swank2016}. The Bloch-Siegert induced frequency shift is generally larger for the co-magnetometer. The magnitude of ${\bf B}_{E \times v}$ for $v = 30\,{\rm m\,s^{-1}}$ and our electric field will be $ \approx 20\,{\rm \mu G}$. This field is transverse to ${\bf B}_0$ with a direction that fluctuates after every collision with the walls or phonons in the superfluid. This results in Bloch-Siegert frequency shifts \cite{Bloch1940}. The combination of this ${\bf B}_{E \times v}$ field with magnetic field gradients transverse to ${\bf B}_0$ produces a frequency shift that is proportional to $E$ and hence will appear as a false EDM signal. The size of this frequency shift is also generally dependent on the gyromagnetic ratio, the static field strength $B_0$, the field gradients, the collisional frequency, and (as already mentioned in the introduction) on the dimensions of the cell. However, closed form expressions exist for the ballistic and highly diffusive cases \cite{Pendlebury2004,Pignol2015}. We will call this the "Bloch-Siegert induced false EDM''. 

A technique to suppress this false EDM effect is to apply a magnetic field gradient and scan its value to interpolate for when the total gradient is zero \cite{Pendlebury2015}. For our $^3$He co-magnetometer, we have another important handle to control this effect. By scanning the temperature of the superfluid helium bath the $^3$He mean-free-path, and thus the collisional frequency, can be changed drastically without influencing the UCNs. For example a $\pm 0.1\,{\rm K}$ scan will cause the collisional frequency to change by a factor of about 50 because of the strong temperature dependence of the phonon density. Combined with previous theoretical treatments of this problem \cite{Lamoreaux2005,Barabanov2006, Golub2011, Swank2012a,Steyerl2014, Golub2015, Pignol2015, Swank2016}, we have a powerful tool for studying and controlling this important false EDM effect. For instance, we can suppress the false EDM by appropriately tuning the temperature, or reduce the temperature temporarily to magnify this effect and use it to tune out field gradients. 

An effect introduced by the polarized $^3$He is the pseudo-magnetic field experienced by the UCNs \cite{Abragam1973}. This comes from the spin-dependence of the real part of the n-$^3$He scattering length \cite{Huber2009,Zimmer2002}. The pseudo-magnetic field experienced by the UCNs will be parallel to the $^3$He spins. For $x_3 = 10^{-10}$ and $P_3 = 1$, the magnitude of this pseudo-magnetic field is $\approx 0.1\,{\rm \mu G}$. The effects of this field can be suppressed by precise operation of the experiment (see Sec.~\ref{sec:SOSapp} about the Systematics and Operational Studies apparatus). 


\subsection{Dressed spin measurement mode} \label{sec:dressedspin}

The second measurement mode is the \emph{dressed spin mode}, where a strong off-resonant AC magnetic field transverse to ${\bf B}_0$ with magnitude $B_d$ and angular frequency $\omega_d$ will be applied \cite{Haroche1970}. In the limit $\gamma B_0  \ll \omega_{d}$, the dressing field causes the gyromagnetic ratio of an unperturbed species $\gamma$ to be modified to:
\begin{equation}
\gamma^\prime \approx \gamma \, J_0\! \left( \frac{\gamma B_{d}}{\omega_{d}} \right) \;,
\label{eq:spindress}
\end{equation}
where $J_0$ is the zeroth-order Bessel function of the first kind. A higher-order treatment of spin dressing and descriptions of the spin motions can be found in \cite{Swank2018a,Tavakoli-Dinani2018}.

By an appropriate choice of dressing field parameters, the effective gyromagnetic ratios of the $^3$He and UCNs can be made the same, i.e. $\gamma_n^\prime = \gamma_3^\prime$. This is first satisfied when $|\gamma_n| B_d/\omega_d \approx 1.19 \equiv x_c$. For $B_{d} =  1\,{\rm G}$, a dressing frequency $f_{d} = \omega_{d}/(2\pi) \approx 2.5 \, {\rm kHz}$ can be used.

Considering for now the case with no $E$ field, by applying the dressing field satisfying $x_c$ the relative azimuthal angle between the two species $\phi_{3n}(t)\equiv \phi_{3}(t) -  \phi_{n}(t) $ can be made to remain fixed. Furthermore, by applying the dressing field slightly above or below $x_c$ (e.g. by shifting $f_d$), then $\phi_{3n}(t)$ can be increased or decreased. 

The pseudo-magnetic field experienced by UCNs when $\phi_{3n}(t) \neq 0$ will cause their spins to precess out of the transverse plane at a frequency of $\approx 300\,{\rm \mu Hz}$ about the $^3$He spin axis. This motion can be mitigated by appropriately modulating the dressing field away from $x_c$ at a modulation frequency $f_m \gg 300\,{\rm \mu Hz}$ (e.g. $f_m \approx 1\,{\rm Hz}$) to make the motion about $\phi_{3n} = 0$ symmetric. To a first order approximation, this will allow the out-of-plane motion during one half of the modulation cycle to be reversed during the next half cycle.

The effect of the $E$ field on the neutron EDM will be an additional angular frequency so that when at critical dressing the evolution of $\phi_{3n}(t)$ will be given by:
\begin{equation}
\phi_{3n}(t) \approx \phi_d \pm \frac{2 J_0(x_c) \,d_n |E|}{\hbar} t \;,
\label{eq:dressspinangle}
\end{equation}
in the $\gamma B_0  \ll \omega_{d}$ limit. Here, $\phi_d$ is some user chosen phase between the neutron and $^3$He spins. 

There are different spin dressing modulation techniques and feedback modes (e.g. for stabilizing the spin dressing field) that can be implemented with different statistics and suppression of systematic drifts. These are described in Refs.~\cite{Golub1994a,Swank2018a,Filippone2019}. With these techniques, the neutron EDM signal (from the additional term in Eq.~\ref{eq:dressspinangle}) can be extracted from a single dressed spin measurement in a single cell. Therefore, the SQUID readout of the $^3$He will not be needed.

The optimized 1$\sigma$ statistical precision from the dressed spin mode with pulsed modulation and using both cells is expected to be ${\sigma_{d_n} \approx 1.6\times 10^{-28}\,e{\rm \,\cdot\, cm}}$ for 300 live days of data taking \cite{Filippone2019}. The same parameters as in Sec.~\ref{sec:freeprecess} are assumed except for $\bar{\tau}_3 = 100\,{\rm s}$, which is more statistically advantageous for this mode. This sensitivity is better than in the free precession mode. Furthermore, this technique will contain different systematics. For instance, there will be no Bloch-Siegert induced false EDM from gradients in the dressing field, and effects from static field inhomogeneities can be suppressed \cite{Swank2018a}. The false EDM due to static field gradients will still be present but can be suppressed with temperature tuning \cite{Swank2018a}, similar to the free precession mode. With the dressed spin mode, precise simultaneous control of the two spin species will be important. The techniques required to achieve this and their associated effects will be studied with the Systematics and Operation Studies apparatus described in Sec.~\ref{sec:SOSapp}. Having these two measurement modes with different systematic effects will allow us to provide an important self-verification of our results.

\section{Apparatus design \label{sec:expDesign}}

\begin{figure}
\centering
\includegraphics[width=\columnwidth,clip]{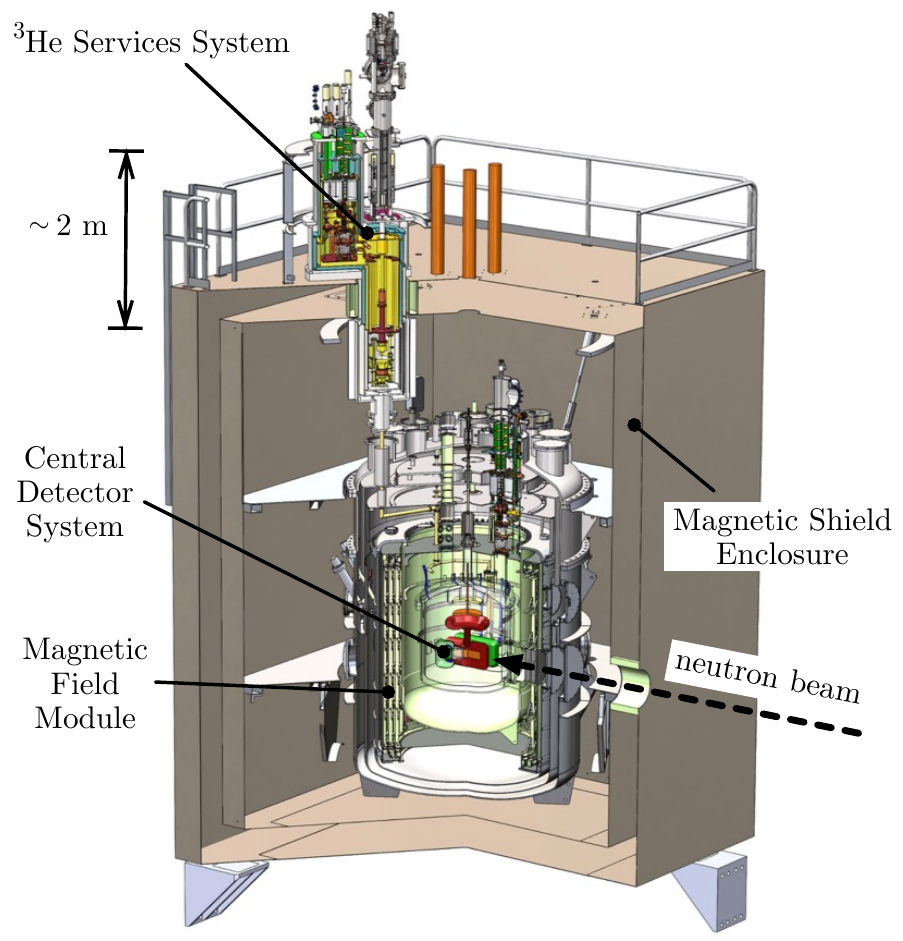}
\caption{Design of the nEDM@SNS apparatus. Primary sub-systems, which are described in Sec.~\ref{sec:expDesign}, are labeled. A magnified view of the Central Detector System is shown in Fig.~(\ref{fig:CDS}).}
\label{fig:wholeExp}
\end{figure}

An overview of our apparatus is shown in Fig.~(\ref{fig:wholeExp}). It consists of three main modular sub-systems: the Central Detector System, the $^3$He Services System, and the Magnetic Field Module. These systems are being developed and tested in parallel at different institutes before their integration at the SNS.  A brief overview of each system, as well as the Systematics and Operational Studies apparatus, is given below. 

\subsection{Central detector system} 

The Central Detector System (CDS), shown in greater detail in Fig.~(\ref{fig:CDS}), will be a $\approx 1600\,{\rm L}$  volume filled with superfluid helium cooled with a custom non-magnetic dilution refrigerator to temperatures in the range $0.3 - 0.5 \,{\rm K}$. This will enable the study of systematic effects as described in Sec.~\ref{sec:3Hecomagnetometer}. The CDS will contain the two measurement cells and associated cell valves, a high voltage electrode, two ground electrodes, a high voltage amplification system, SQUID arrays, and scintillation light detection systems.

Materials inside the CDS volume have stringent requirements in terms of being non-magnetic (to avoid producing field gradients) and non-electrically conductive to avoid Johnson noise in the SQUIDs and eddy current heating caused by the spin dressing field. The volume itself will be constructed from an epoxy-impregnated fiberglass composite material similar to G10. Large scale ($\approx 50\,{\rm cm}$ diameter) detachable cryogenic seals made using a Kapton gasket placed between G10 plates have been developed for the seal of the CDS volume. Cryogenic stress-relieved epoxy mounts for large silicon wafers, required for cold neutron beam windows, have also been developed.

The measurement cells are required to have low UCN reflection wall loss probability and low $^3$He and UCN wall depolarization rates. They must be $^3$He leak tight (to exclude $^3$He which exists at natural isotopic abundance levels in the $1600\,{\rm L}$ superfluid $^4$He of the CDS), and be constructed from high purity and deuterated materials where needed.

Full-sized cells made from protonated PMMA plates with the dTPB-doped dPS (dPS+dTPB) coating on the inside have been produced. The neutron optical potential of the coatings have been verified to be $\approx 160\,{\rm neV}$ with neutron reflectometry. The UCN storage properties of these cells have been characterized using the LANL solid deuterium based UCN source \cite{Saunders2013,Ito2018} and an apparatus that can cool the cells down to $\approx 20\,{\rm K}$ in vacuum. A UCN wall loss time $\tau_{\rm walls}\approx 2000\,{\rm s}$ has been demonstrated for 50\% of UCNs loaded into the cell from this source with stainless steel guides. The dPS+dTPB coating has also been demonstrated to have low depolarization probability per $^3$He wall collision ($\approx 1 \times 10^{-7}$) down to $0.3\,{\rm K}$ \cite{Ye2009,Yoder2010} in small ($\approx 20\,{\rm cm^3}$) cells.

A valve is needed for loading and unloading a cell with $^3$He (see Sec.~\ref{sec:3HeS}). This valve when closed is required to be UCN and $^3$He storage friendly and UCN and $^3$He polarization friendly, similar to the cells. A prototype deuterated plastic cell valve system has also been tested in the cryogenic UCN storage test apparatus. This valve has demonstrated UCN loss at levels well below the wall loss rate. The $^3$He holding time at this valve has also been demonstrated to be sufficiently long from the results of extrapolating helium gas holding measurements performed down to $4\,{\rm K}$. 

In another apparatus designed for studying electrical breakdown in $0.4\,{\rm K}$ superfluid helium \cite{Ito2016}, a $12\,{\rm cm}$ diameter copper implanted PMMA electrode has been demonstrated to support a stable electric field $> 75\,{\rm kV/cm}$. Being a stochastic process, the dielectric breakdown probability scales with the electrode surface area. The leakage currents measured for PMMA in various geometries inserted in between the electrodes were demonstrated to be low enough to reduce leakage current systematic effects to $<1.5 \times 10^{-28}\, e{\rm \,\cdot\, cm}$ for pessimistic assumptions.

In order to reach the $75\,{\rm kV/cm}$ design $E$ field inside the cells, a potential of $\approx 630\,{\rm kV}$ is required at the high voltage electrode. This is too high for a direct feed system. A suitable high voltage amplification system has been identified in the form of a cryogenic Cavallo multiplier system. Here, charge is transferred to the high voltage electrode in discrete steps by a moving electrode. A room temperature prototype has been used to demonstrate this technique \cite{Clayton2018}.

After a ${\rm n\textrm{-}^3He}$ capture event approximately $4,600$ EUV photons with wavelengths $\approx 80\,{\rm nm}$ are expected to be produced in the superfluid helium in the presence of the $E$ field \cite{Ito2012}. These photons need to be detected with sufficiently high efficiency to enable discrimination against backgrounds (from neutron $\beta$ decay and other sources). The scintillation light detection efficiency has been measured in small ($\approx 100\,{\rm cm^3}$) prototype cells coated with dPS+dTPB. These cells were filled with $4\,{\rm K}$ liquid helium and scintillation light was produced with a $^{210}$Po $\alpha$-source. The light collection system employed wavelength shifting fibers to convert blue light from the TPB to green light. This green light was then guided to silicon photomultipliers (SiPM). Using the results from these tests for the design of a similar system in the nEDM@SNS apparatus, $\approx 20$ photoelectrons are expected to be detected by the SiPMs per ${\rm n + ^3He}$ event. These results were used to determine the values of $\epsilon_3$ and $\epsilon_\beta$ described in Sec.~\ref{sec:spinanalysis}.

SQUIDs will be used to detect the nuclear precession signal of the $^3$He co-magnetometer. The intrinsic and flux noise of a prototype SQUID system with long pick-up coil leads has been demonstrated to meet the noise requirements at $4\,{\rm K}$ \cite{Kim2013}.
 
\subsection{$^3$He services system \label{sec:3HeS}}

The cryostat of the $^3$He Services (He3S) System, located several meters above the CDS, will provide the temperature gradient (relative to the measurement cell temperature) for loading polarized $^3$He before each measurement cycle and unloading (partially) depolarized $^3$He afterwards. This $^3$He ``heat flush'' injection and removal technique is described in \cite{Baym2015} for $^3$He concentrations and temperatures relevant to our proposed experiments. Experimental heat flush tests have been performed using superfluid helium with $^3$He at concentrations of $x_3\approx 10^{-6}$.

The polarized $^3$He source will be a cryocooler based atomic beam, which was previously used in the experiments described in \cite{Esler2007,Eckel2012}. With a nozzle cooled to $\approx 1\,{\rm K}$, a cold effusive source of $^3$He atoms is produced. This unpolarized $^3$He passes through a permanent magnetic quadrupole that filters out one spin-state, producing a polarized $^3$He beam with a flux of $\approx 10^{14}\,{\rm atoms/s}$. This beam will be injected onto the surface of a small superfluid helium filled volume (called the ``Injection Volume'') in the He3S cryostat. A spin transport magnetic field system will be used to tailor the magnetic field so as to maintain the $^3$He polarization during injection. Scattering of the $^3$He beam by $^4$He vapor arising from evaporation of the superfluid film from the Injection Volume will be suppressed by a film burner that has been fabricated and successfully tested.

After accumulation, the polarized $^3$He in the Injection Volume will be transported to the cell. Polarized $^3$He friendly coatings for the transfer tube to the cell have been developed \cite{Yoder2010}. The $^3$He removed from the cell after each measurement will be concentrated by heat flushing to a small volume. This volume, called the ``Sequestration Volume,'' will then be isolated, and the liquid inside evaporated and removed with room temperature pumps.

A separate dilution refrigerator will be used for cooling the He3S cryostat. Since this will be located further away from the cells than the refrigerator for the CDS, its non-magnetic requirements are slightly relaxed. This dilution refrigerator is currently under construction and has been cooled to $1\,{\rm K}$, demonstrating superfluid leak tightness of the components.

\subsection{Magnetic field module}

The Magnetic Field Module (MFM), which surrounds the CDS, will provide the $30\,{\rm mG}$ static $B_0$ field as well as the AC fields (transverse to ${\bf B}_0$) for the $\pi/2$ rotation and spin dressing. Field gradients cause spin relaxation and give rise to the Bloch-Siegert induced false EDM effect, as discussed in Secs.~\ref{sec:freeprecess} and \ref{sec:3Hecomagnetometer}. The magnetic field gradient requirements for the $B_0$ field are less than a few ${\rm ppm/cm}$, while those for the spin dressing field are less stringent by a factor of about 15.

To produce the uniform horizontal magnetic fields, modified saddle-shaped cos$\,\theta$ coil designs will be used. In order from smallest radius outwards, the components of the MFM (all of which have a vertical cylindrical geometry) are: the spin dressing coil, an eddy current shield (a thin conductor to reduce heating of external material caused by the spin dressing field), the $B_0$ coil, a ferromagnetic shield (which provides a flux return for the $B_0$ field, improving its uniformity inside the coil), a superconducting Pb magnetic shield (for suppression of time varying fields), and a magnetic cloak (comprising  strips of ferromagnetic material to produce a tunable effective permeability, described more in Sec.~\ref{sec:MSE}). The Pb magnetic shield is closed at the top and bottom with Pb end caps, and with the curved saddle wires of the cos$\,\theta$ coils located outside. The end caps also serve as magnetic mirrors to increase the effective length of the cos$\,\theta$ coils, improving field uniformity.

The cryostat for the MFM will have an outer vacuum jacket, a liquid nitrogen cooled shield, and an inner magnet volume (IMV) for cooling the components described above. The outer shell of the IMV will be cooled with liquid helium lines, and inside the IMV will be helium gas to provide heat exchange. This will facilitate operation of the superconducting Pb shield, as well as superconducting Pb alloy coil wires at temperatures below $7\,{\rm K}$. The inner shell of the IMV is required to be made from a non-electrically conductive material because of eddy current heating and Johnson noise requirements.

The construction and testing of a half scale prototype MFM system has enabled us to demonstrate our ability to produce the required field gradients, maintain adequate shielding, and sustain necessary heat loads \cite{Galvan2011,Slutsky2017}. The outer aluminum vacuum vessel and liquid nitrogen shields for the final MFM have been fabricated and cryogenically tested. 

An array of fluxgate magnetometers will be located between the inner wall of the MFM and the CDS \cite{Nouri2014,Nouri2014a,Nouri2015}. This magnetometer array will be used to reconstruct the field at the two measurement cells by solving Laplace's equation for the magnetic scalar potential ($\Phi_{\rm mag}$) assuming no current  or magnetization sources inside the boundary of the magnetometer array (i.e. ${\bf \nabla}^2\Phi_{\rm mag}=0$). This system will be used for initial tuning of magnetic field gradients.

\subsection{Magnetic shield enclosure \label{sec:MSE}}

A multi-layered ferromagnetic Magnetic Shield Enclosure (MSE) with internal dimensions approximately $4\times 4 \times 6\,{\rm m^3}$ will be located outside the MFM. A magnetically sealed door will provide internal access. Specifications for the MSE include field gradients $\lesssim 1\,{\rm nT/m}$ over a $1\,{\rm m^3}$ volume at the cells, and dynamic shielding factors $>150$ at $1\,{\rm Hz}$ and $> 10^{4}$ between $0.1$ and $1\,{\rm kHz}$. A tri-axial system of rectangular coils will surround the MSE to compensate for the Earth's field and other background fields.

A set of coils that produce a uniform field parallel to ${\bf B}_0$ will be located inside the MSE. This uniform field will help maintain $^3$He polarization during transport from the He3S System. It will also help reduce field distortions caused by gaps in the superconducting Pb shield end caps of the MFM. Along the side wall of the Pb shield, the magnetic field of the internal MSE coils will encounter the flux repelling superconductor. The ferromagnetic cloak inside the MFM (mentioned earlier) will be used to cancel the effects of the superconductor in order to match the field inside and outside the Pb shield.

\subsection{Cold neutron beam}

The cold neutron beam will pass through the MSE and the MFM, and enter the CDS to reach the superfluid helium inside the measurement cells in order to produce UCNs. The beam line has been redesigned to incorporate a super-mirror for polarization and a chopper system for monochromation to $8.9\,{\textrm{\AA}}$, rather than the intercalated graphite Bragg-reflection monochrometers initially proposed \cite{Fomin2015}. Despite the fact that the revised neutronics system will require longer guides, better optimization of angles and beam cross-sections will improve the integrated $8.9\,{\textrm{\AA}}$ cold neutron flux by $\sim 40\%$. This will provide a volumetric UCN production rate in the cells of $\approx 0.31\,{\rm UCN\,cm^{-3}\,s^{-1}}$ for our trapping potential ($160\,{\rm neV}$ for dPS minus $18.5\,{\rm neV}$ for the superfluid helium).


\subsection{Systematics and operational studies apparatus \label{sec:SOSapp}}

The Systematics and Operational Studies (SOS) apparatus will be used to perform measurements required for finalizing the design of various components and protocols of the nEDM@SNS experiment. It will be a smaller apparatus with a cryogenic turnaround time of approximately two weeks, as compared to two months for the main apparatus.

The design of the SOS apparatus does not include a high voltage system and also contains only one measurement cell. This reduces the volume of superfluid helium that is required from $\approx 1600\,{\rm L}$ to $\approx 5\,{\rm L}$. Polarized $^3$He will be provided by a room-temperature metastability exchange optical pumping (MEOP) based source. This source enables $^3$He to be loaded to concentrations that are several orders of magnitude higher compared with the atomic beam source, at the expense of a lower $P_3$. These higher concentrations will be useful for certain measurements that don't require long UCN storage times. The UCNs will be injected from an external source rather than being produced in situ. The magnetic field gradient requirements in the SOS apparatus are also relaxed by a factor of three relative to the main apparatus. These design specifications make the SOS apparatus easier to implement.

The key measurements that will be performed with the SOS apparatus will include:
\begin{enumerate}
\item Characterizing the UCN storage properties, $^3$He and UCN wall spin relaxation times, and fluorescence properties of the measurement cells prior to installation in the nEDM@SNS apparatus.
\item Characterizing UCN and $^3$He motional correlation functions which are essentially unaffected by the $E$-field, and which can then be used to validate and control the Bloch-Siegert induced false EDM.
\item Identifying techniques required for the high-precision simultaneous control of the two spin species, such as those required for the $\pi/2$ pulse and spin dressing.
\end{enumerate}
Note in regard to the latter that spin dressing of $^3$He in an atomic beam \cite{Esler2007,Eckel2012} and in a cell \cite{Chu2011a} has been previously demonstrated. Similarly, critical spin dressing on $^1$H and $^{19}$F in a liquid sample has also been performed \cite{Tavakoli-Dinani2018}. However, the duration and precision of control in these experiments are several orders of magnitude below what is required for the nEDM@SNS experiment.


The SOS apparatus will enable us to divert key studies from the critical path of the main nEDM@SNS experiment and will decrease the time needed to reach a physics result. The design of the SOS apparatus and the associated measurement program will be described in an upcoming paper \cite{Leung2018}.

The aluminum cryostat needed for the SOS apparatus has been constructed and commissioned to $4\,{\rm K}$. The MEOP source achieved the design goal of $P_3 = 0.8$ inside the optical pumping cell. The dilution refrigerator that will be used has been cooled and observed to deliver the cooling power required. Various demanding cryogenic components have been fabricated and tested, such as superfluid leak tight cryogenic components and seals made only from plastics, and components for the $^3$He removal system.

\section{Conclusion}

The nEDM@SNS experiment will search for a neutron EDM down to the $\mathcal{O}(10^{-28}\, e{\rm \,\cdot\, cm})$ level which is compelling for tests of electroweak baryogenesis and beyond standard model physics. This unprecedented sensitivity will be reached using a high UCN density produced directly in superfluid helium inside the measurement cell, as well as by performing in situ and live UCN spin analysis using polarized $^3$He. The relatively small measurement cell size combined with a unique cryogenic $^3$He co-magnetometer allows many key systematic effects to be controlled and sufficiently suppressed, in particular the Bloch-Siegert induced false EDM effect. The apparatus allows two distinct measurement techniques with different systematics: the free precession mode and the dressed spin mode. These two modes of operation exhibit different sensitivities to effects giving rise to statistical and systematic uncertainties. This will provide the opportunity for a powerful self-check of our results. The combined systematic error of our experiment is expected to be below $ 2 \times 10^{-28}\, e{\rm \,\cdot\, cm}$.

The nEDM@SNS collaboration has devoted significant effort to the development of the apparatus and techniques for the realization of this pioneering experiment proposed by \cite{Golub1994a}. Over the last five years we have experimentally demonstrated technical readiness of many high fidelity prototypes and standalone final components. The valuable lessons learned have been applied to an engineering design that we believe will reduce schedule risks. In mid-2018, we transitioned to a ``large scale integration'' phase during which we will make large-scale procurements, construct final components, and integrate them at the SNS. The acquisition of first physics data is expected to begin toward the end of 2023. The apparatus is expected to reach $\mathcal{O}(10^{-27}\, e{\rm \,\cdot\, cm})$ statistical precision after only 1 week of measurement time.

\section{Acknowledgements}

This work was supported in part by the U.S. Department of Energy under grants DE-AC02-06CH11357, DE-AC05-00OR22725, DE-AC52-06NA25396, DE-FG02-94ER40818, DE-FG02-97ER41042, DE-FG02-99ER41101, DE-SC0014622, DE-SC0008107, DE-SC0005367, DE-FG02-88ER40416, and 2017LANLEEDM, and the U.S. National Science Foundation under grants 1306547, 1205977, 1506459, 1812340, 1440011, 1506416, 1822502, 1812377, and 1506451.

\bibliography{PPNS2018_Leung}

\begin{thebibliography}{76}

\bibitem{Schwinger1951}
J.~Schwinger, Phys. Rev. \textbf{82}, 914 (1951)

\bibitem{Luders1957}
G.~L{\"u}ders, Ann. Phys. (N. Y.) \textbf{2}, 1  (1957)

\bibitem{Christenson1964}
J.H. Christenson, J.W. Cronin, V.L. Fitch, R.~Turlay, Phys. Rev. Lett.
  \textbf{13}, 138 (1964)

\bibitem{Aubert2002}
B.~Aubert, D.~Boutigny, J.M. Gaillard, A.~Hicheur, Y.~Karyotakis, J.P. Lees,
  P.~Robbe, V.~Tisserand, A.~Zghiche, A.~Palano et~al. (The \emph{BABAR}
  Collaboration), Phys. Rev. Lett. \textbf{89}, 201802 (2002)

\bibitem{Abe2002}
K.~Abe, K.~Abe, T.~Abe, I.~Adachi, H.~Aihara, M.~Akatsu, Y.~Asano, T.~Aso,
  V.~Aulchenko, T.~Aushev et~al. (Belle Collaboration), Phys. Rev. D
  \textbf{66}, 071102 (2002)

\bibitem{Cabibbo1963}
N.~Cabibbo, Phys. Rev. Lett. \textbf{10}, 531 (1963)

\bibitem{Kobayashi1973}
M.~Kobayashi, T.~Maskawa, Progr. Theor. Exp. Phys. \textbf{49}, 652 (1973)

\bibitem{Sakharov1967}
A.D. Sakharov, Pisma Zh. Eksp. Teor. Fiz. \textbf{5}, 32 (1967), [JETP Lett. 5
  (1967) 24] [Sov. Phys. Usp. 34 (1991) 392]

\bibitem{Li2009}
Y.~Li, S.~Profumo, M.~Ramsey-Musolf, Phys. Lett. B \textbf{673}, 95  (2009)

\bibitem{Cirigliano2010}
V.~Cirigliano, Y.~Li, S.~Profumo, M.J. Ramsey-Musolf, J. High Energy Phys.
  \textbf{2010}, 2 (2010)

\bibitem{Ramsey1950}
N.F. Ramsey, Phys. Rev. \textbf{78}, 695 (1950)

\bibitem{Smith1957}
J.H. Smith, E.M. Purcell, N.F. Ramsey, Phys. Rev. \textbf{108} (1957)

\bibitem{Ramsey1982}
N.F. Ramsey, Rep. Prog. Phys. \textbf{45}, 95 (1982)

\bibitem{Dress1977}
W.B. Dress, P.D. Miller, J.M. Pendlebury, P.~Perrin, N.F. Ramsey, Phys. Rev. D
  \textbf{15}, 9 (1977)

\bibitem{Lamoreaux2009}
S.K. Lamoreaux, R.~Golub, J. Phys. G \textbf{36}, 104002 (2009)

\bibitem{Baker2006}
C.A. Baker, D.D. Doyle, P.~Geltenbort, K.~Green, M.G.D. van~der Grinten, P.G.
  Harris, P.~Iaydjiev, S.N. Ivanov, D.J.R. May, J.M. Pendlebury et~al., Phys.
  Rev. Lett. \textbf{97}, 131801 (2006)

\bibitem{Pendlebury2015}
J.M. Pendlebury, S.~Afach, N.J. Ayres, C.A. Baker, G.~Ban, G.~Bison, K.~Bodek,
  M.~Burghoff, P.~Geltenbort, K.~Green et~al., Phys. Rev. D \textbf{92}, 092003
  (2015)

\bibitem{Baker2014}
C.~Baker, Y.~Chibane, M.~Chouder, P.~Geltenbort, K.~Green, P.~Harris,
  B.~Heckel, P.~Iaydjiev, S.~Ivanov, I.~Kilvington et~al., Nucl. Instrum.
  Methods Phys. Res. A \textbf{736}, 184  (2014)

\bibitem{Anghel2009}
A.~Anghel, F.~Atchison, B.~Blau, B.~van~den Brandt, M.~Daum, R.~Doelling,
  M.~Dubs, P.A. Duperrex, A.~Fuchs, D.~George et~al., Nucl. Instrum. Methods
  Phys. Res. A \textbf{611}, 272  (2009)

\bibitem{Becker2015}
H.~Becker, G.~Bison, B.~Blau, Z.~Chowdhuri, J.~Eikenberg, M.~Fertl, K.~Kirch,
  B.~Lauss, G.~Perret, D.~Reggiani et~al., Nucl. Instrum. Methods Phys. Res. A
  \textbf{777}, 20 (2015)

\bibitem{Roccia2018}
{S. Roccia et al. (PSI nEDM collaboration)}, These proceedings (2018)

\bibitem{Commins1991}
E.D. Commins, Am. J. Phys. \textbf{59}, 1077 (1991)

\bibitem{Pendlebury2004}
J.M. Pendlebury, W.~Heil, Y.~Sobolev, P.G. Harris, J.D. Richardson, R.J.
  Baskin, D.D. Doyle, P.~Geltenbort, K.~Green, M.G.D. van~der Grinten et~al.,
  Phys. Rev. A \textbf{70}, 032102 (2004)

\bibitem{McGregor1990}
D.D. McGregor, Phys. Rev. A \textbf{41}, 2631 (1990)

\bibitem{Golub1994a}
R.~Golub, S.K. Lamoreaux, Phys. Rep. \textbf{237}, 1 (1994)

\bibitem{Filippone2019}
M.W. Ahmed et~al., arxiv:1908.09937  (2019)

\bibitem{Fomin2015}
N.~Fomin, G.~Greene, R.~Allen, V.~Cianciolo, C.~Crawford, T.M. Ito, P.~Huffman,
  E.~Iverson, R.~Mahurin, W.~Snow, Nucl. Instrum. Methods Phys. Res. A
  \textbf{773}, 45  (2015)

\bibitem{Golub1975}
R.~Golub, J.M. Pendlebury, Phys. Lett. A \textbf{53}, 133 (1975)

\bibitem{Golub1977}
R.~Golub, J.M. Pendlebury, Phys. Lett. A \textbf{62}, 337 (1977)

\bibitem{Golub1979}
R.~Golub, Phys. Lett. A \textbf{72}, 387 (1979)

\bibitem{Golub1983}
R.~Golub, C.~Jewell, P.~Ageron, W.~Mampe, B.~Heckel, I.~Kilvington, Z. Phys. B
  \textbf{51}, 187 (1983)

\bibitem{Yoshiki1992}
H.~Yoshiki, K.~Sakai, M.~Ogura, T.~Kawai, Y.~Masuda, T.~Nakajima, T.~Takayama,
  S.~Tanaka, A.~Yamaguchi, Phys. Rev. Lett. \textbf{68}, 1323 (1992)

\bibitem{Baker2003}
C.A. Baker, S.N. Balashov, J.~Butterworth, P.~Geltenbort, K.~Green, P.G.
  Harris, M.G.D. van~der Grinten, P.S. Iaydjiev, S.N. Ivanov, J.M. Pendlebury
  et~al., Phys. Lett. A \textbf{308}, 67 (2003)

\bibitem{Zimmer2007a}
O.~Zimmer, K.~Baumann, M.~Fertl, B.~Franke, S.~Mironov, C.~Plonka, D.~Rich,
  P.~Schmidt-Wellenburg, H.F. Wirth, B.~van~den Brandt, Phys. Rev. Lett.
  \textbf{99}, 104801 (2007)

\bibitem{Piegsa2014}
F.M. Piegsa, M.~Fertl, S.N. Ivanov, M.~Kreuz, K.K.H. Leung,
  P.~Schmidt-Wellenburg, T.~Soldner, O.~Zimmer, Phys. Rev. C \textbf{90},
  015501 (2014)

\bibitem{Leung2016}
K.K.H. Leung, S.~Ivanov, F.M. Piegsa, M.~Simson, O.~Zimmer, Phys. Rev. C
  \textbf{93}, 025501 (2016)

\bibitem{Gerhold1998}
J.~Gerhold, Cryogenics \textbf{38}, 1063  (1998)

\bibitem{Ito2016}
T.M. Ito, J.C. Ramsey, W.~Yao, D.H. Beck, V.~Cianciolo, S.M. Clayton,
  C.~Crawford, S.A. Currie, B.W. Filippone, W.C. Griffith et~al., Rev. Sci.
  Instrum. \textbf{87}, 045113 (2016)

\bibitem{Passell1966}
L.~Passell, R.I. Schermer, Phys. Rev. \textbf{150}, 146 (1966)

\bibitem{Ito2012}
T.M. Ito, S.M. Clayton, J.~Ramsey, M.~Karcz, C.Y. Liu, J.C. Long, T.G. Reddy,
  G.M. Seidel, Phys. Rev. A \textbf{85}, 042718 (2012)

\bibitem{Ito2013}
T.M. Ito, G.M. Seidel, Phys. Rev. C \textbf{88}, 025805 (2013)

\bibitem{Archibald2006}
G.~Archibald, J.~Boissevain, R.~Golub, C.R. Gould, M.E. Hayden, E.~Korobkina,
  W.S. Wilburn, J.~Zou, AIP Conf. Proc. \textbf{850}, 143 (2006)

\bibitem{Schiff1963}
L.I. Schiff, Phys. Rev. \textbf{132}, 2194 (1963)

\bibitem{Flambaum2012}
V.V. Flambaum, A.~Kozlov, Phys. Rev. C \textbf{85}, 068502 (2012)

\bibitem{Kim2013}
Y.J. Kim, S.M. Clayton, IEEE Trans. Appl. Supercond. \textbf{23}, 2500104
  (2013)

\bibitem{Baym2013}
G.~Baym, D.H. Beck, C.J. Pethick, Phys. Rev. B \textbf{88}, 014512 (2013)

\bibitem{Lamoreaux2005}
S.K. Lamoreaux, R.~Golub, Phys. Rev. A \textbf{71}, 032104 (2005)

\bibitem{Barabanov2006}
A.L. Barabanov, R.~Golub, S.K. Lamoreaux, Phys. Rev. A \textbf{74}, 052115
  (2006)

\bibitem{Golub2011}
R.~Golub, R.M. Rohm, C.M. Swank, Phys. Rev. A \textbf{83}, 023402 (2011)

\bibitem{Swank2012a}
C.~Swank, A.~Petukhov, R.~Golub, Phys. Lett. A \textbf{376}, 2319  (2012)

\bibitem{Steyerl2014}
A.~Steyerl, C.~Kaufman, G.~M\"uller, S.S. Malik, A.M. Desai, R.~Golub, Phys.
  Rev. A \textbf{89}, 052129 (2014)

\bibitem{Golub2015}
R.~Golub, C.~Kaufman, G.~M\"uller, A.~Steyerl, Phys. Rev. A \textbf{92}, 062123
  (2015)

\bibitem{Pignol2015}
G.~Pignol, M.~Guigue, A.~Petukhov, R.~Golub, Phys. Rev. A \textbf{92}, 053407
  (2015)

\bibitem{Swank2016}
C.M. Swank, A.K. Petukhov, R.~Golub, Phys. Rev. A \textbf{93}, 062703 (2016)

\bibitem{Bloch1940}
F.~Bloch, A.~Siegert, Phys. Rev. \textbf{57}, 522 (1940)

\bibitem{Abragam1973}
A.~Abragam, G.L. Bacchella, H.~Gl\"atti, P.~Meriel, M.~Pinot, J.~Piesvaux,
  Phys. Rev. Lett. \textbf{31}, 776 (1973)

\bibitem{Huber2009}
M.G. Huber, M.~Arif, T.C. Black, W.C. Chen, T.R. Gentile, D.S. Hussey, D.A.
  Pushin, F.E. Wietfeldt, L.~Yang, Phys. Rev. Lett. \textbf{102}, 200401 (2009)

\bibitem{Zimmer2002}
O.~Zimmer, G.~Ehlers, B.~Farago, H.~Humblot, W.K. nd~R.~Scherm, EPJ direct
  \textbf{4}, 1 (2002)

\bibitem{Haroche1970}
S.~Haroche, C.~Cohen-Tannoudji, C.~Audoin, J.P. Schermann, Phys. Rev. Lett.
  \textbf{24}, 861 (1970)

\bibitem{Swank2018a}
C.M. Swank, E.K. Webb, X.~Liu, B.W. Filippone, Phys. Rev. A \textbf{98}, 053414
  (2018)

\bibitem{Tavakoli-Dinani2018}
R.~Tavakoli~Dinani, Ph.D. thesis, Simon Fraser University (2018)

\bibitem{Saunders2013}
A.~Saunders, M.~Makela, Y.~Bagdasarova, H.O. Back, J.~Boissevain, L.J.
  Broussard, T.J. Bowles, R.~Carr, S.A. Currie, B.~Filippone et~al., Rev. Sci.
  Instrum. \textbf{84}, 013304 (2013)

\bibitem{Ito2018}
T.M. Ito, E.R. Adamek, N.B. Callahan, J.H. Choi, S.M. Clayton, C.~Cude-Woods,
  S.~Currie, X.~Ding, D.E. Fellers, P.~Geltenbort et~al., Phys. Rev. C
  \textbf{97}, 012501 (2018)

\bibitem{Ye2009}
Q.~Ye, H.~Gao, W.~Zheng, D.~Dutta, F.~Dubose, R.~Golub, P.~Huffman, C.~Swank,
  E.~Korobkina, Phys. Rev. A \textbf{80}, 023403 (2009)

\bibitem{Yoder2010}
J.~Yoder, Ph.D. thesis, University of Illinois at Urbana-Champaign (2010)

\bibitem{Clayton2018}
S.~Clayton, T.~Ito, J.~Ramsey, W.~Wei, M.~Blatnik, B.~Filippone, G.~Seidel, J.
  Instrum. \textbf{13}, P05017 (2018)

\bibitem{Baym2015}
G.~Baym, D.H. Beck, C.J. Pethick, Phys. Rev. B \textbf{92}, 024504 (2015)

\bibitem{Esler2007}
A.~Esler, J.C. Peng, D.~Chandler, D.~Howell, S.K. Lamoreaux, C.Y. Liu, J.R.
  Torgerson, Phys. Rev. C \textbf{76}, 051302 (2007)

\bibitem{Eckel2012}
S.~Eckel, S.K. Lamoreaux, M.E. Hayden, T.M. Ito, Phys. Rev. A \textbf{85},
  032124 (2012)

\bibitem{Galvan2011}
A.~{P{\'e}rez Galv{\'a}n}, B.~Plaster, J.~Boissevain, R.~Carr, B.~Filippone,
  M.~Mendenhall, R.~Schmid, R.~Alarcon, S.~Balascuta, Nucl. Instrum. Methods
  Phys. Res. A \textbf{660}, 147  (2011)

\bibitem{Slutsky2017}
S.~Slutsky, C.~Swank, A.~Biswas, R.~Carr, J.~Escribano, B.~Filippone,
  W.~Griffith, M.~Mendenhall, N.~Nouri, C.~Osthelder et~al., Nucl. Instrum.
  Methods Phys. Res. A \textbf{862}, 36  (2017)

\bibitem{Nouri2014}
N.~Nouri, B.~Plaster, J. Instrum. \textbf{9}, P11009 (2014)

\bibitem{Nouri2014a}
N.~Nouri, B.~Plaster, Nucl. Instrum. Methods Phys. Res. A \textbf{767}, 92
  (2014)

\bibitem{Nouri2015}
N.~Nouri, A.~Biswas, M.~Brown, R.~Carr, B.~Filippone, C.~Osthelder, B.~Plaster,
  S.~Slutsky, C.~Swank, J. Instrum. \textbf{10}, P12003 (2015)

\bibitem{Chu2011a}
P.H. Chu, Phys. Rev. C \textbf{84}, 022501 (2011)

\bibitem{Leung2018}
K.K.H. Leung et~al., (in preparation) (2019)

\end{thebibliography}

\end{document}